\documentclass[letterpaper, 10 pt, conference]{ieeeconf}  

\IEEEoverridecommandlockouts                   
\usepackage[bookmarks=false]{hyperref}
\usepackage{graphics} 
\usepackage{epsfig} 
\usepackage{mathptmx} 
\usepackage{times} 
\usepackage{amsmath} 
\usepackage{amssymb}  
\usepackage[T1]{fontenc}
\usepackage[latin9]{inputenc}
\usepackage{amsmath}
\usepackage{graphicx}
\usepackage{authblk}
\usepackage{hyperref}
\usepackage{color}

\newcommand{\revise}[1]{\textcolor{black}{#1}}

\title{\LARGE \bf Isolating the impact of trading on grid frequency fluctuations}

\author{Benjamin Sch\"afer$^{1}$, Marc Timme$^{1}$ and Dirk Witthaut$^{2}$
\thanks{
This work is supported through the German Science Foundation (DFG)
by a grant toward the Cluster of Excellence \textquotedblleft Center
for Advancing Electronics Dresden\textquotedblright{} (cfaed), the
Helmholtz Association (via the joint initiative \textquotedblleft Energy
System 2050\textemdash a Contribution of the Research Field Energy\textquotedblright{}
and grant no. VH-NG-1025) and the Federal Ministry of Education and
Research (BMBF Grant No 03SF0472F).}
\thanks{$^{1}$Benjamin Sch\"afer and Marc Timme  are with  Chair for Network Dynamics, Center for Advancing Electronics Dresden (cfaed), Institute for Theoretical Physics, Technical University of Dresden, Dresden and Network Dynamics, MPIDS, G\"ottingen, Germany      
        {\tt\small benjamin.schaefer@tu-dresden.de | marc.timme@tu-dresden.de}}%
\thanks{$^{2}$Dirk Witthaut is with the Forschungszentrum J\"ulich, Institute of Energy and Climate Research-Systems Analysis and Technology Evaluation, J\"ulich and Institute for Theoretical Physics, University of Cologne, K\"oln, Germany
        {\tt\small d.witthaut@fz-juelich.de}}%
        \thanks{\\978-1-5386-4505-5/18/\$31.00 ©2018 IEEE}
}

\begin{document}

\maketitle
\thispagestyle{empty}
\pagestyle{empty}

\begin{abstract}
To ensure reliable operation of  power grids, their frequency shall stay within strict bounds. Multiple sources of disturbances cause fluctuations of the grid frequency, ranging from changing demand over volatile feed-in to energy trading. Here, we analyze frequency time series from the continental European grid in 2011 and 2017 as a case study to isolate the impact of trading. 
We find that trading at typical trading intervals such as full hours modifies the frequency fluctuation statistics. While particularly large frequency deviations in 2017 are not as frequent as in 2011, large deviations are more likely to occur shortly after the trading instances.
A comparison between the two years indicates that trading at shorter intervals might be beneficial for frequency quality and grid stability, because particularly large fluctuations are substantially diminished.
Furthermore, we observe that the statistics of the frequency fluctuations do not follow Gaussian distributions but are better described using heavy-tailed and asymmetric distributions, for example L{\'e}vy-stable distributions.
Comparing intervals without trading to those with trading instances indicates that frequency deviations near the trading times are distributed more widely and thus extreme deviations are orders of magnitude more likely.
Finally, we briefly review a stochastic analysis that allows a quantitative description of power grid frequency fluctuations.
\end{abstract}

\section{Introduction}
Power generation and consumption have to be balanced to allow the power grid to operate close to its reference frequency (e.g.,
$f_{R}=50~\text{Hz}$ in Europe) and thereby ensure robust distribution among generators and consumers \cite{Machowski2011}.
However, the grid frequency does not stay at precisely $f_{R}=50~\text{Hz}$ during operation but is subject to multiple sources of power fluctuations, ranging from demand fluctuations \cite{Wood2013} over stochastic feed-in by renewable sources \cite{Sorensen2007,Milan2013} to effects caused by energy trading \cite{NationalAcademiesofSciences2016}. 
Fluctuations of the grid frequency must not exceed certain security  limits 
\cite{Kundur1994,Omran2011}. 
 Whereas renewable energy sources undoubtedly challenge the system due to their distributed and variable nature \cite{Boemer2011,Rohden2012,Wohland2018}, 
an analysis on the German system has shown that renewables only put light stress on primary control reserves \cite{Hirth2015}.  
 In contrast, impacts due to energy trading seem to be substantial, as discussed in  \cite{Weisbach2009,ENTSO-E2011} and recently in \cite{Schaefer2017a}.

When setting up a smart grid \cite{Amin2005,Fang2012,Schaefer2015,Schaefer2016}, it is crucial to know the underlying systemic dynamics and potential vulnerabilities of the system.
Liberalizing the energy market \cite{NationalAcademiesofSciences2016} may have economic upsides yet it is not fully clear to date how it impacts a grid's stability \cite{Schaefer2017b}.
Including active consumers, e.g., via demand control schemes \cite{Albadi2008,Martyr2018}, will also influence the frequency statistics. So one key question is: Can we isolate effects of trading in power grid frequency recordings and quantify their impact?

Here, we analyze frequency data from Continental Europe from 2011 and 2017 as a case study. We first observe the impact of trading based on hourly mean frequency trajectories and investigate the aggregated distribution of frequency values. To isolate effects by trading, we split the given data set into time windows surrounding the 15-minutes trading intervals. Evaluating measures such as standard deviations and kurtosis, we quantify differences between trading and non-trading time intervals of the time series.
Finally, we briefly review important stochastic results on quantifying frequency distributions \cite{Schaefer2017a}.

\section{Analyzing frequency time series}
\begin{figure}
\begin{centering}
\includegraphics[width=0.95\columnwidth]{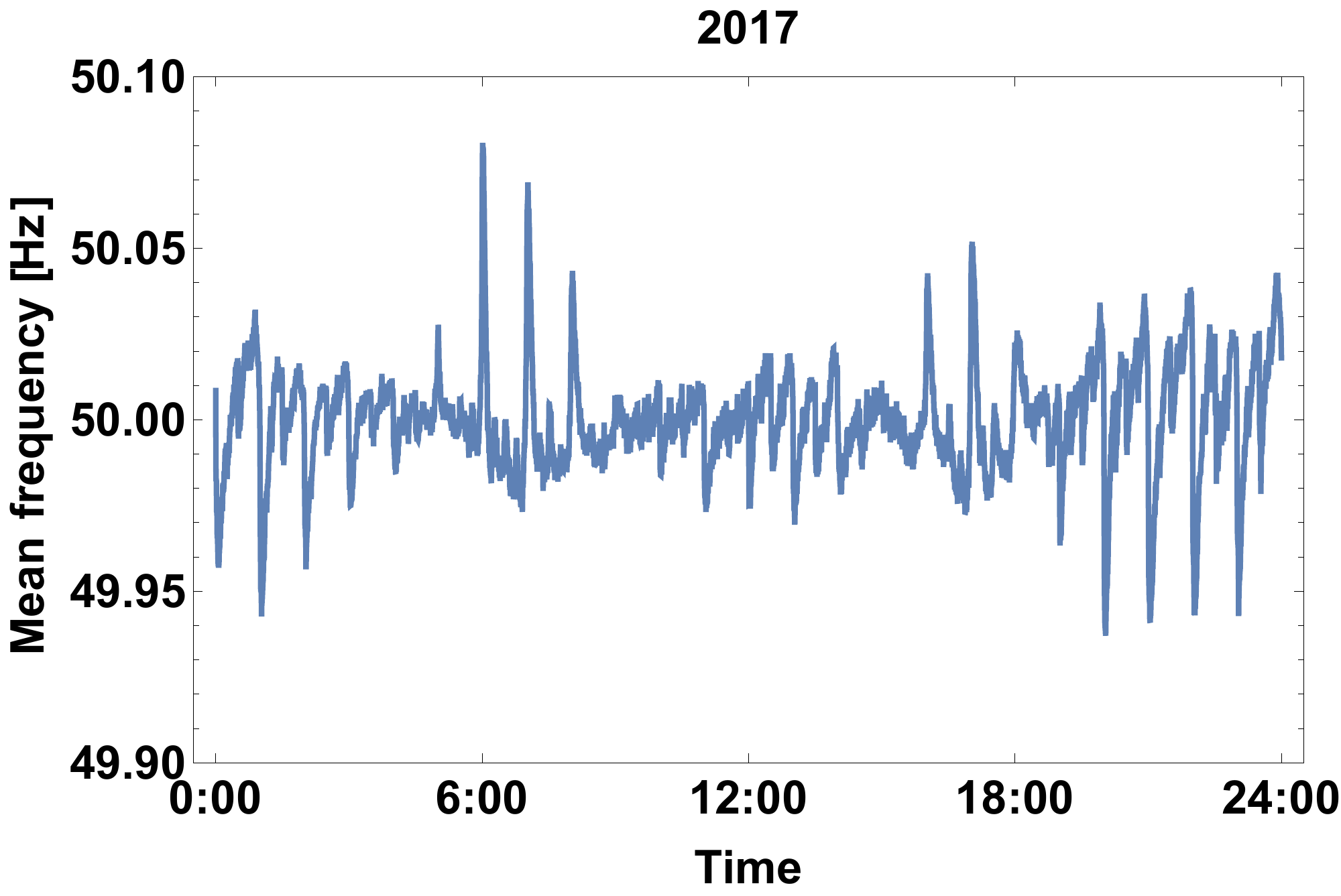}
\par\end{centering}
\caption{The mean frequency of the 2017 time series is mostly close to the reference frequency of $f_R=50~\text{Hz}$ but displays noticeable peaks at fixed intervals, in particular at full hours. Plotted is the mean frequency across every day, based on the \emph{RTE} data set \cite{RTE-UCTE2016}.
\label{fig:meanFrequency2017}}
\end{figure}

\begin{figure}
\begin{centering}
\includegraphics[width=0.95\columnwidth]{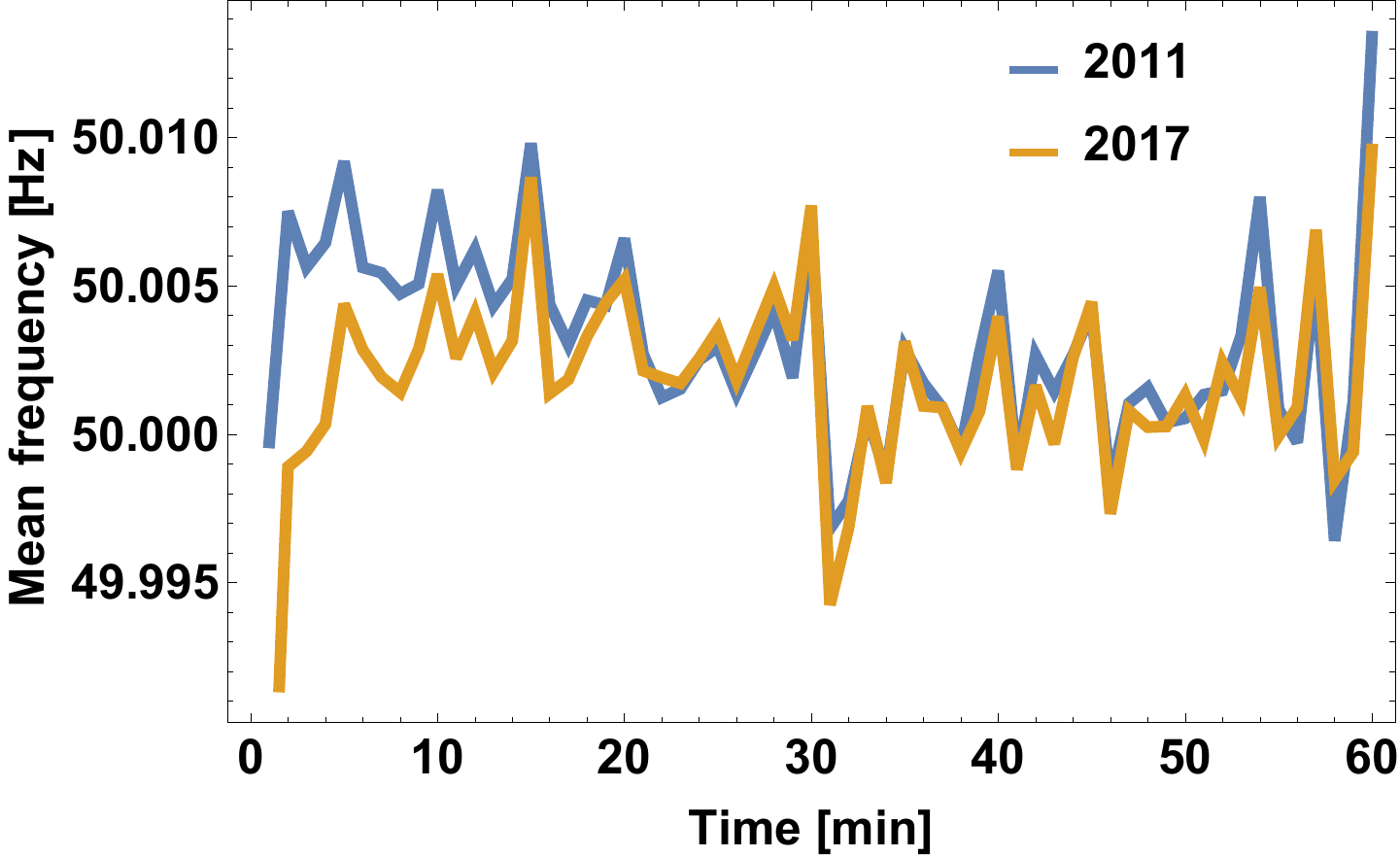}
\par\end{centering}
\caption{The mean hourly frequency of both the 2011 and 2017 data sets displays large deviations and jumps at the beginning and end of every hour (minutes 0 and 60) as well as approximately 30 minutes after full hours. We plot the mean frequency for all hours considered in our 2011 \cite{50Hertz-UCTE2016} and 2017 \cite{RTE-UCTE2016} data set.
\label{fig:meanHourlyFrequency}}
\end{figure}

\begin{figure}
\begin{centering}
\includegraphics[width=0.95\columnwidth]{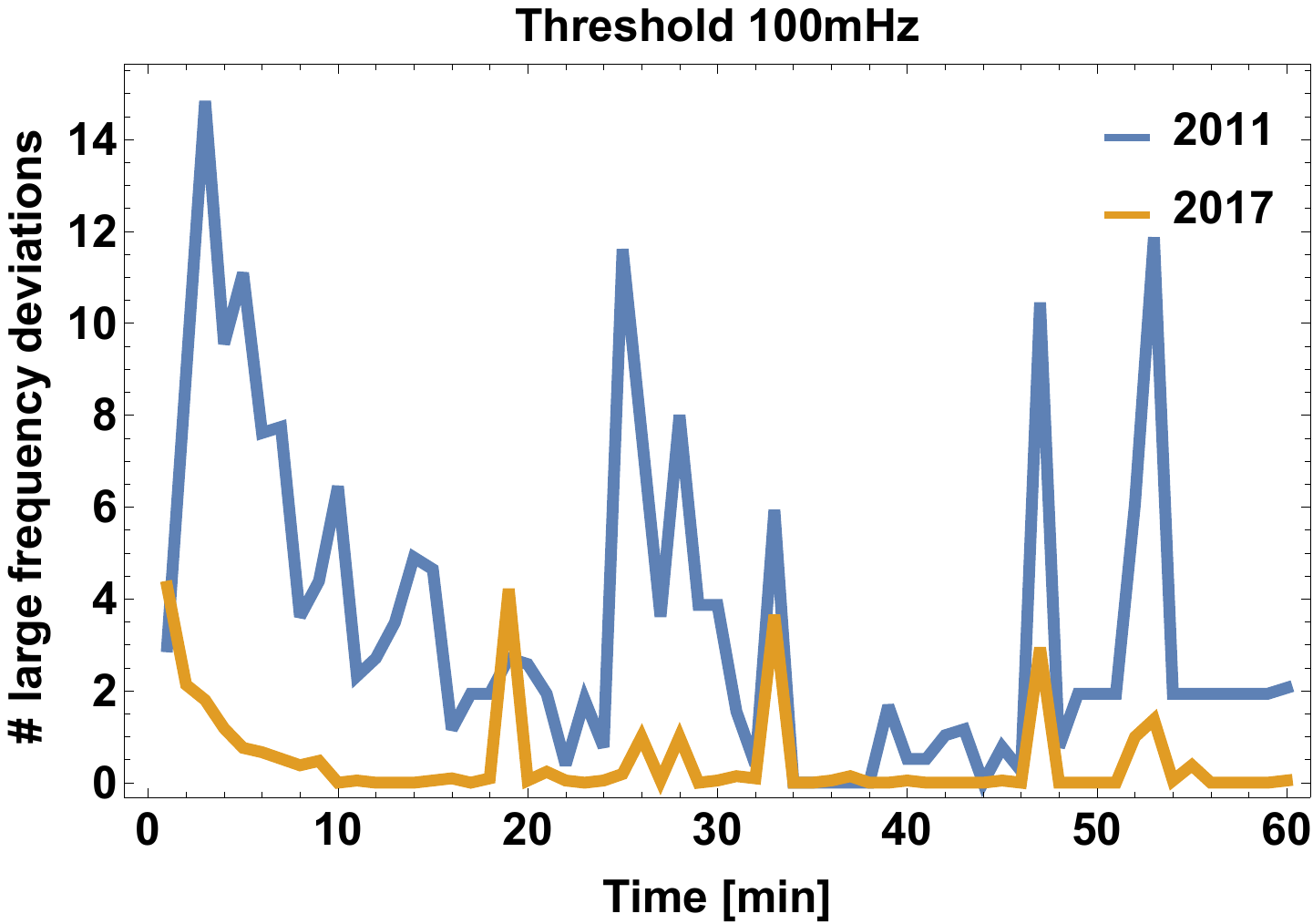}
\par\end{centering}
\caption{Large deviations from the reference frequency were more likely in 2011. Recently, large deviations are likely to occur a few minutes after the trading actions every 15 minutes.
We plot the mean number of measurement points within one hour for which the frequency deviations surpass a threshold of $100~m\text{Hz}$.
To this end, we aggregated all data into hour-blocks, as in Fig.~\ref{fig:meanHourlyFrequency}. Within each hour, we counted on a minute-by-minute basis how often the frequency deviates from the reference of $50~\text{Hz}$ by more than the threshold of $100~m\text{Hz}$. We normalized for the number of days and the different time resolution in the two data sets, assuming a 1 second resolution. 
\label{fig:thresholdViolations}}
\end{figure}

\begin{figure}
\begin{centering}
\includegraphics[width=0.95\columnwidth]{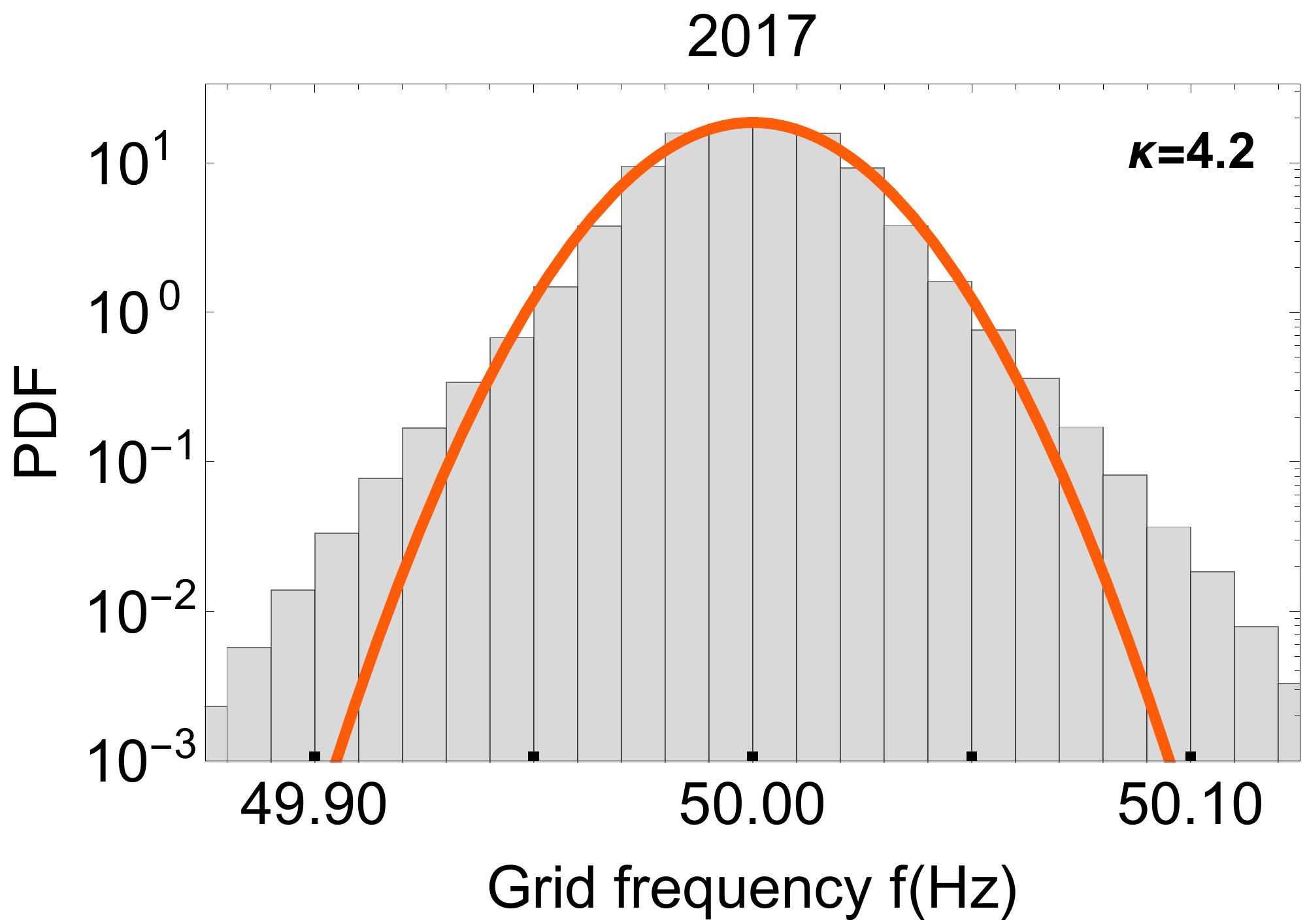}
\par\end{centering}
\caption{The overall frequency distribution is not well described by a Gaussian distribution (red line) but has heavier tails. Shown is the histogram of the aggregated 2017 data \cite{RTE-UCTE2016}. The Gaussian fit was obtained using maximum a likelihood analysis. The data displays a kurtosis of $\kappa\approx 4.2$ in contrast to the Gaussian $\kappa^\text{Gauss}=3$.
\label{fig:HistogramsOfOriginalData}}
\end{figure}

We analyze frequency statistics using recordings provided by \emph{Réseau de Transport d'Electricité (RTE)} \cite{RTE-UCTE2016} and \emph{50Hertz} \cite{50Hertz-UCTE2016}, describing the Continental European synchronous zone. \emph{RTE} provides very recent data so that we analyze the year 2017, while \emph{50Hertz} offers also older data so that we analyze time series from 2011. The \emph{RTE} data has a resolution of one measurement every 10 seconds while the \emph{50Hertz} data has one measurement every 4 seconds. Unfortunately, both data sets contain gaps or invalid values. \revise{Hence, we restrict our analysis for 2011 to January for a total of 31 days and for 2017 we include January, February, May, August, September, November and December (consisting only of 30 days of measurement) for a total of 211 days of complete data which we use without further processing.}

For both 2011 and 2017, the mean frequency per day is computed, i.e., all measurements taken at a specific time are averaged over all available days, e.g. all measurements taken at 0:00 on all days are averaged to obtain the mean frequency at 0:00. We observe that the mean frequency is generally close to the reference frequency of $f_R=50~\text{Hz}$ but shows regular large deviations throughout the day (Fig.~\ref{fig:meanFrequency2017}). \revise{An earlier study  \cite{Weisbach2009} found that deviations tend to be towards negative frequencies during the night and noon and towards positive values during morning and afternoon hours and are more pronounced every full hour, which is consistent with our observations. The direction of the jumps could be determined by the ramps of the demand curve. 
To highlight the jumps, we aggregate the full data into one hour blocks and average recordings, e.g. measurements at 0:00, 1:00, etc. are averaged to obtain the hourly mean for 0 min. Analyzing the result, we indeed observe regular spikes of the mean frequency at one hour and a smaller one at 30 minutes (Fig.~\ref{fig:meanHourlyFrequency}).}


\revise{To further explore how larger frequency deviations are temporally distributed, we record every instance when the frequency deviation  $\Delta f=|f-f_R|$ surpassed a threshold of $f_\text{Threshold}=100~m\text{Hz}$. To do so, we aggregate the data into hour-blocks and analyze them on a minute-by-minute basis. 2011 showed more of these large frequency deviations, while recently the violations mainly occur shortly after the trading interval every 15 minutes (Fig \ref{fig:thresholdViolations}).}

Next, we investigate the distribution of the frequency values, aggregated over the 2017 data set in Fig.~\ref{fig:HistogramsOfOriginalData}. We notice that the best Gaussian fit substantially underestimates the tails of the distribution. We quantify these heavy tails by displaying the kurtosis of the grid, which is approximately $\kappa \approx 4.2$, compared to the Gaussian $\kappa^\text{Gauss}=3$. This means that large disturbances are much more likely than expected based on a Gaussian assumption.
In particular, the fraction of instances, where $\Delta f>f_\text{Threshold}$, increases approximately by a factor of 160.
 Therefore, non-Gaussian distributions, for example L{\'e}vy-stable \cite{Samorodnitsky1994} or q-Gaussian \cite{Tsallis2009}, should rather be considered to model the frequency distribution \cite{Schaefer2017a}.

\section{Impacts of trading}
\begin{figure}
\begin{centering}
\includegraphics[width=0.9\columnwidth]{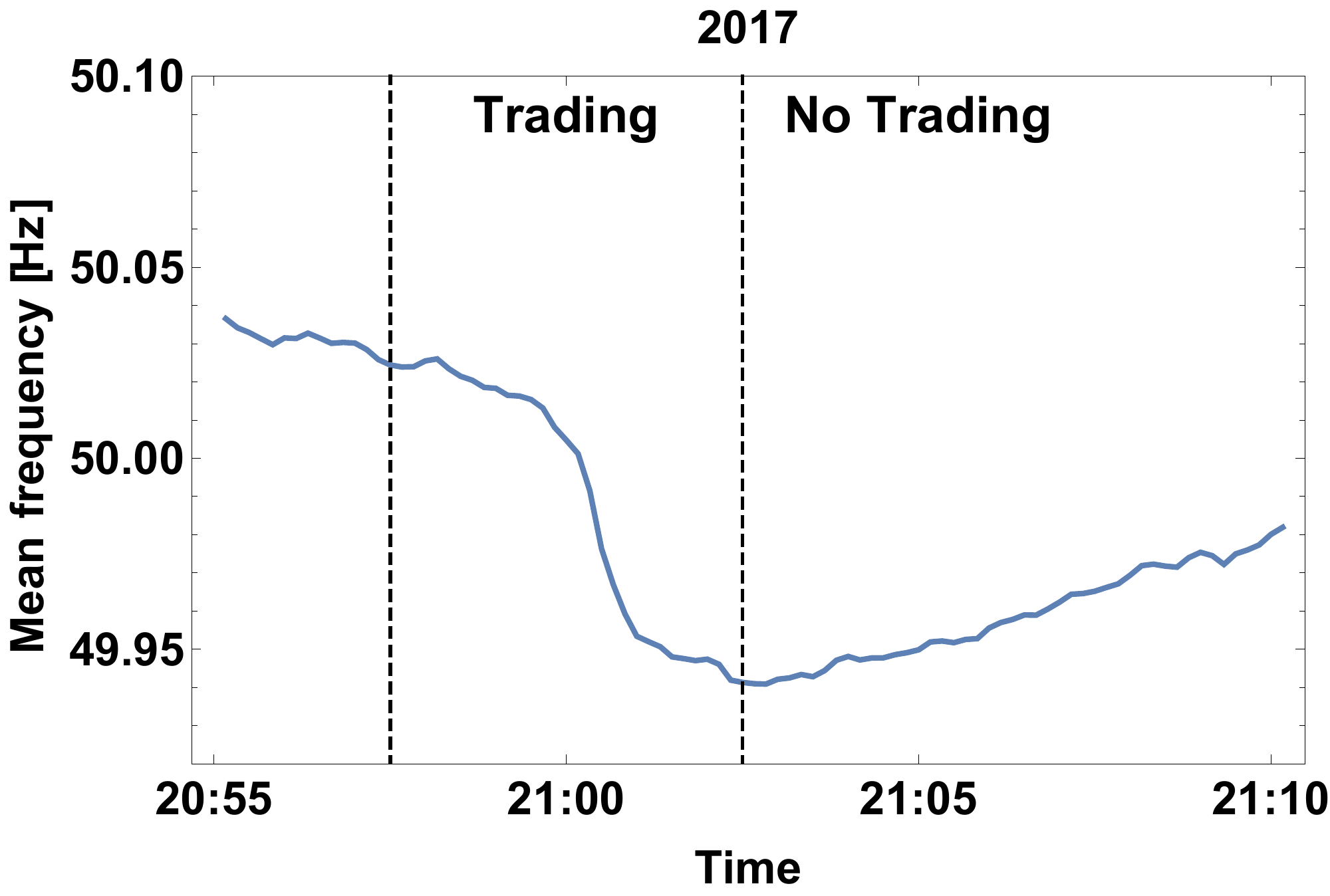}
\par\end{centering}
\caption{Illustration of how we extract the trading effects: Around every 15 minute interval, we extract a 5 minute frequency window of available frequency measurements to islolate the impacts of trading activities on frequency quality. Displayed is the mean frequency of the 2017 data \cite{RTE-UCTE2016}.}
\label{fig:Illustation-Trading-Interval}
\end{figure}

\begin{figure*}
\begin{centering}
\includegraphics[width=1.9\columnwidth]{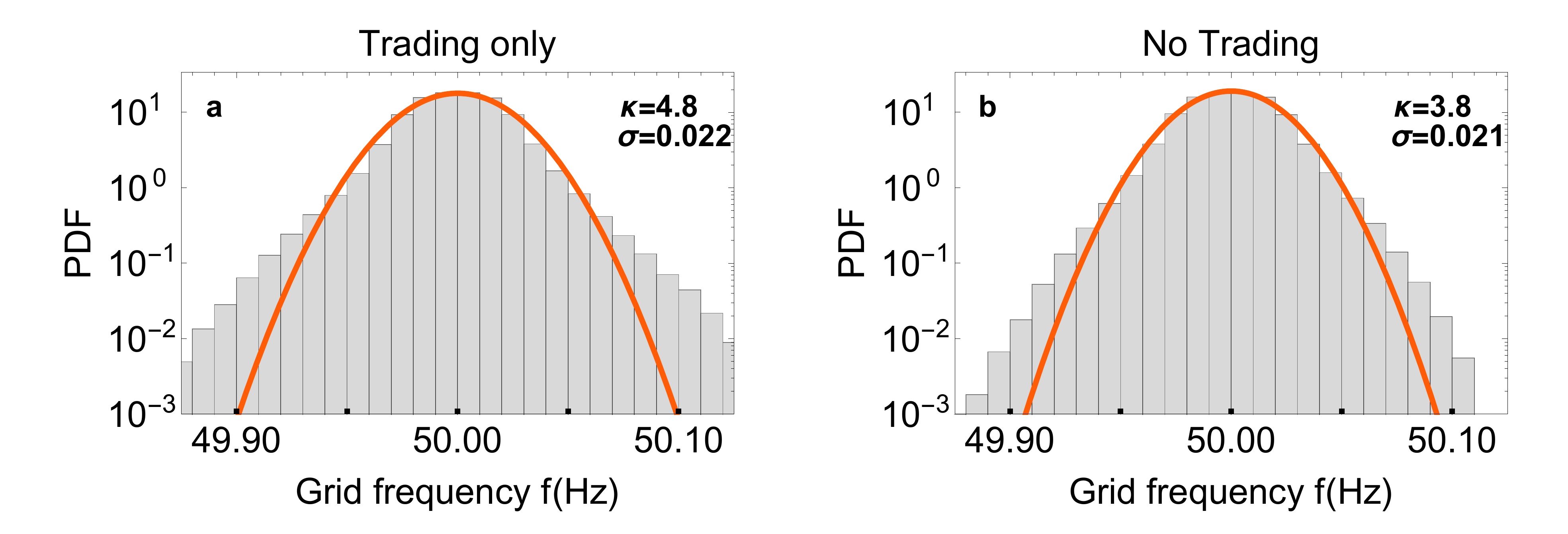}
\par\end{centering}
\caption{
The frequency distributions at the trading interval is broader and displays heavier tails than the data outside the trading intervals.
a: Histogram of the aggregated 2017 data, using only data within a trading window (5 minutes centered around the 15 minute trading intervals). b: Histogram of the aggregated 2017 data, using only data outside of the trading window.
The kurtosis of the data taken at the trading interval is much larger than the kurtosis outside the trading intervals. In contrast, the standard deviation between the data sets differs only by  approximately $5-10\%$.
\label{fig:TradingVsNonTradingHisto}}
\end{figure*}

Does trading significantly impact the width and the heavy tails of the distribution? Following the observation that the frequency
displays jumps predominantly at the trading intervals every 15 minutes (see Figs.~\ref{fig:meanFrequency2017} and \cite{Schaefer2017a}), we isolate
those time periods and compare histograms only consisting of data
around the trading intervals with the remaining distributions for all other intervals without trading. To this end we assume
that the trading takes place every full hour, at 15 minutes, 30 minutes
and 45 minutes after every full hour. Around those time stamps, we cut a window of 5 minutes
of the frequency recordings to be analyzed as the frequency at the
trading interval while the remaining data is left for comparison. See Fig.~\ref{fig:Illustation-Trading-Interval} for an illustration, using frequency recordings from \emph{RTE} \cite{RTE-UCTE2016}.

\revise{Trading causes the distribution to be only slightly wider, yet implies substantially more pronounced heavy tails (Fig.~\ref{fig:TradingVsNonTradingHisto}). 
Similar to the full data set, we observe significant deviations from Gaussianity, especially for the trading time windows (panel a), but also during non-trading intervals (panel b). To quantify the difference between the two data sets, we compute the width of the distributions, measured via sample standard deviation and the strength of the heavy-tails of the distributions, quantified via the kurtosis in Fig.~\ref{fig:TradingVsNonTradingHisto}. While the standard deviation only differs by less than $10\%$, the trading interval data is much more heavy-tailed ($\kappa=4.8$) than the non-trading interval ($\kappa=3.8$), implying a high chance for large deviations (extreme events).}

\section{A stochastic model for power grids}
We observed non-Gaussian statistics for power grid frequency fluctuations in the histograms of the aggregated data (Figs. \ref{fig:HistogramsOfOriginalData} and \ref{fig:TradingVsNonTradingHisto}).
To anticipate the risks due to large frequency excursions, we require a stochastic theory that links power fluctuations with frequency fluctuations. This would allow us to anticipate the magnitude of fluctuations when designing a new system, such as a microgrid or smart grid. We will briefly review some corner stones of the results recently obtained in \cite{Schaefer2017a} on this question.

First, we model the power grid as a network of (virtual) synchronous machines \cite{Biegel2014}. A simple dynamical description for such a network applies the swing equation  \cite{Machowski2011,Kundur1994}, which gives the dynamics of the voltage phase angle $\theta_{i}\left(t\right)$ and the angular velocity
$\omega_{i}\left(t\right)$ at each node $i\in\left\{ 1,...,N\right\} $ as
\begin{eqnarray}
\frac{\text{d}}{\text{d}t}\theta_{i} & = & \omega_{i},\label{eq:Equation of motion power grid}\\
M_{i}\frac{\text{d}}{\text{d}t}\omega_{i}  & = & P_{i}+\Gamma_i(t)-D_{i}\omega_{i}+\sum_{j=1}^{N}K_{ij}\sin\left(\theta_{j}-\theta_{i}\right),\nonumber 
\end{eqnarray}
with inertia $M_{i}$, active power $P_{i}$,
random fluctuations $\Gamma_i(t)$, damping $D_{i}$ and coupling matrix $K_{ij}$. To convert from frequencies to angular velocities one uses the following conversion
\begin{equation}
\omega=2\pi \left(f-f_R\right).
\end{equation}

To allow further analysis of the grid dynamics, we simplify this system by assuming symmetric coupling $K_{ij}=K_{ji}$, homogeneous damping to inertia ratio \cite{Motter2013a} $\gamma=D_{i}/M_{i}$ and balanced power \emph{on average} $\sum_{i=1}^{N}P_{i}=0$. Then, the dynamics of the bulk angular velocity
$\bar{\omega}=\sum_{i=1}^{N}\omega_{i}M_{i}/\sum_{i=1}^{N}M_{i}$
is given by 

\begin{equation}
\frac{\text{d}}{\text{d}t}\bar{\omega}=-\gamma\bar{\omega}+\frac{\sum_{i=1}^N\Gamma_i(t)}{\sum_{i=1}^NM_i}.\label{eq:bulk dynamics}
\end{equation} 

Assuming the noise $\Gamma_i$ at each node $i$ is following a Gaussian distribution with standard deviation $\sigma^P_i$, then we may apply a Fokker-Planck equation\cite{Schaefer2017a,Gardiner1985,Risken1984a} to compute the probability distribution of the angular velocity $\omega$. It is also a Gaussian distribution with standard deviation

\begin{equation}
\bar{\sigma}^{\omega}=\frac{1}{\sum_{i=1}^{N}M_{i}}\sqrt{\frac{\sum_{i=1}^{N}\left(\sigma_{i}^{P}\right)^{2}}{2\gamma}}.\label{eq:Predicted standard deviation}
\end{equation}

To estimate the damping to inertia ration $\gamma$ in Eq. \ref{eq:Predicted standard deviation}, we take advantage of the autocorrelation function of the frequency signal. Assuming approximately Gaussian noise, stochastic theory \cite{Gardiner1985} predicts the autocorrelation $c\left(\Delta t\right)$ to decay as a function of the time delay $\Delta t$ as an exponential function like 
\begin{equation}
c\left(\Delta t\right)=\exp\left(-\gamma\Delta t\right).\label{eq:Autocorrelation decay}
\end{equation}

Finally, to capture non-Gaussian effects of the statistics, as displayed, e.g. in Fig.~\ref{fig:HistogramsOfOriginalData}, we use for example L{\'e}vy-stable distributions \cite{Samorodnitsky1994} as discussed in \cite{Schaefer2017a}. These distributions are characterized by their scale parameter $\sigma_S$, similar to a standard deviation, and their stability parameter $\alpha_S$ that gives the tails of the distribution. Stable distributions with $\alpha_S=2$ are Gaussian distributions while $\alpha_S<2$ indicates heavy tails. For the full 2017 data, we found that $\alpha_S \approx 1.9$, i.e. well below $\alpha_S=2$.
Analogue to the Gaussian approach, we may now formulate a generalized Fokker-Planck  equation \cite{Schaefer2017a,Denisov2009} to obtain the probability distribution of the angular velocity $\omega$.
We require that the noise at each node in the system follows a stable distribution with one stability parameter $\alpha_{S}$ but arbitrary scale parameter, which we set as $\sqrt{2} \sigma_{S,i}^{P}$, where the square root of 2 is necessary to resemble Gaussian results for $\alpha_S=2$. Then, the resulting distribution of the angular velocity is also a stable distribution, with stability parameter $\alpha_S$ and scale parameter given as \cite{Schaefer2017a}
\begin{equation}
\bar{\sigma}_{S}^{\omega}=\frac{1}{\sqrt{2}\sum_{i=1}^{N}M_{i}}\left[\frac{1}{\gamma\alpha_{S}}\sum_{i=1}^{N}\left(\sigma_{S,i}^{P}\right)^{\alpha_{S}}\right]^{1/\alpha_{S}}.\label{eq:Predicted scale parameter}
\end{equation}

\revise{One important finding of equations \eqref{eq:Predicted standard deviation} and \eqref{eq:Predicted scale parameter} is the scaling with respect to the size of the grid. Both the scale parameter $\bar{\sigma}_{S}^{\omega}$  and the standard deviation $\bar{\sigma}^{\omega}$ increase with decreasing total inertia $\sum_i M_i$. Hence, splitting large synchronous grids into smaller ones or replacing conventional generators by inverter technology is expected to increase frequency deviations, unless the primary control, which determines $\gamma$, is also increased.}

How does the previous analysis of trading intervals influence these stochastic results?
The analytical approach in \cite{Schaefer2017a} does not include effects of trading at all but assumes random and uncorrelated noise to account for all frequency disturbances. One important finding of Fig.~\ref{fig:TradingVsNonTradingHisto} is that the width of the distribution, i.e. the standard deviation or scale parameter of the probability distribution does not change dramatically when comparing trading and non-trading intervals. Only the heavy-tails are easily attributed to the trading. 
Hence, the stochastic theory should be further developed to include non-Gaussian effects due to deterministic trading actions, instead of solely arising from random fluctuations.
However, the current form of the theory will still be able to predict the approximate width of a distribution if the noise amplitudes in the system are known. 

\section{Discussion}
Overall, we found that the power grid frequency is impacted by trading actions, as visible already from the average frequency trajectory in Fig.~\ref{fig:meanFrequency2017}.
In particular, the  mean frequency shows consistent jumps every hour and less pronounced effects every 30 minutes. 
When investigating large deviations from the reference frequency we found that the occurrence of these deviations has been reduced when comparing data from 2011 with data from 2017.
Between those years, additional short-time trading products have been established in Europe \cite{NationalAcademiesofSciences2016} so that a larger number of smaller trading actions seems to be beneficial for the frequency quality.

Analyzing the full frequency distribution, we found that it is not well described by a Gaussian distribution but displays heavy tails.  Therefore, more sophisticated distributions like L{\'e}vy-stable \cite{Samorodnitsky1994} or q-Gaussian distributions \cite{Tsallis2009} should be considered when describing frequency distributions. 
Choosing a small time interval around the trading actions every 15 minutes allowed us to reveal that trading has only little impact on the width (standard deviation) of the distribution but much larger impact on the heavy-tails. Nevertheless, even outside trading intervals, the frequency is not well-described by a Gaussian distribution.
A stochastic theory that neglects the effects of trading will therefore overestimate the heavy tails of the noise as those are significantly impacted by trading actions. However, such a theory will give a good approximation for the width of the distribution since the standard deviation does not depend as strongly on the trading actions.
We briefly reviewed some core results obtained in \cite{Schaefer2017a} that allow to quantify the impact of power fluctuations on the power grid frequency even in non-Gaussian settings. 

\revise{While we based our analysis on the equations of motion of synchronoues machines, our results should also hold true if conventional generators were replaced by devices with small or only virtual inertia \cite{Beck2007}, as the equations of motions do not change.}

\revise{Our current analysis emphasizes that future markets have to consider their impact on frequency quality. Based on the comparison between 2011 and 2017, splitting trading actions into smaller packages may reduce the overall impact of trading on the frequency quality, especially with respect to large deviations. This will have to be taken into account in addition to network transmission constraints to balance the system. Otherwise, additional and costly control actions will be necessary \cite{Weisbach2009,ENTSO-E2011}.}

\revise{Future research should explore models for grids without any inertia and include more systematic studies of different grids, e.g. the US grids with different trading intervals or even real time pricing would be very interesting to analyze. However, frequency data are rarely available publicly.}

\bibliographystyle{IEEEtran}
\bibliography{References}


%
%

%

\vspace{-1cm}
\begin{biography}[{\includegraphics[width=1in,height=1.25in,clip,keepaspectratio]{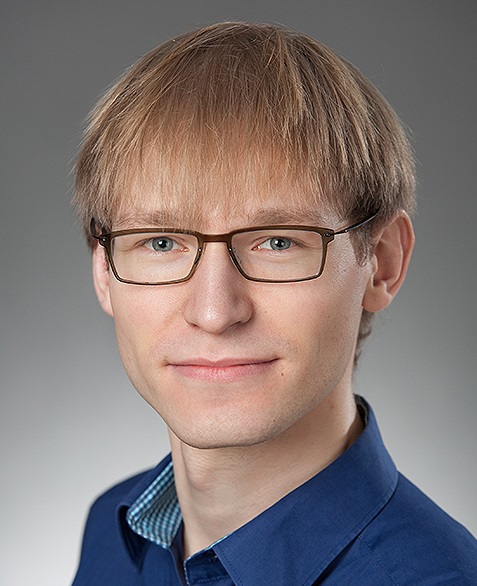}}] {Benjamin Sch\"afer}  received his Diploma (MSc) degree in Physics from the University of Magdeburg, Germany in 2013. Persuing his Ph.D. in G\"ottingen (Germany), London (United Kingdom) and Tokyo (Japan), he received his Ph.D. degree in physics in 2017 from the University of G\"ottingen. Since 2017 he is Postdoctoral Researcher at the Max Planck Institute for Dynamics and Self-Organization, G\"ottingen, Germany and the Technical University Dresden, Germany.  \end{biography}
\vspace{-1cm}
\begin{biography}[{\includegraphics[width=1in,height=1.25in,clip,keepaspectratio]{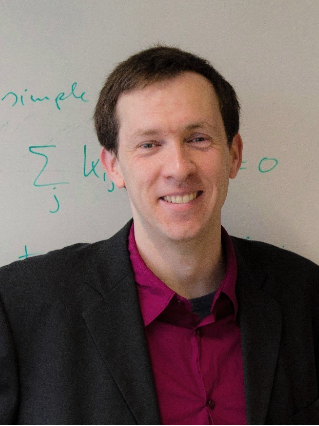}}] {Marc Timme} 
studied Physics and Mathematics at the University of W\"urzburg, Germany, and the State University of New York at Stony Brook, USA. He gained his doctorate in Theoretical Physics at the University of G\"ottingen. After research stays at the Max Planck Institute for Flow Research and the Center for Applied Mathematics, Cornell University (USA), he established and lead a broadly cross-disciplinary research group on Network Dynamics at the Max Planck Institute for Dynamics and Self-Organization, became an Adjunct Professor at the University of G\"ottingen, was Visiting Professor at TU Darmstadt, visiting faculty at the
ETH Zurich Risk Center (Switzerland) 
and recently assumed a Chair for Network Dynamics as a Strategic Professor of the Cluster of Excellence Center for Advancing Electronics Dresden (cfaed) at TU Dresden, Germany.
Marc Timme is Co-Chair of the Division of Socio-Economic Physics of the German Physical Society and received a research award of the Berliner Ungewitter Foundation, the Otto Hahn Medal of the Max Planck Society, and a Research Fellow position of the National Research Center of Italy. 
\end{biography}
%
\vspace{-1cm}
\begin{biography}[{\includegraphics[width=1in,height=1.25in,clip,keepaspectratio]{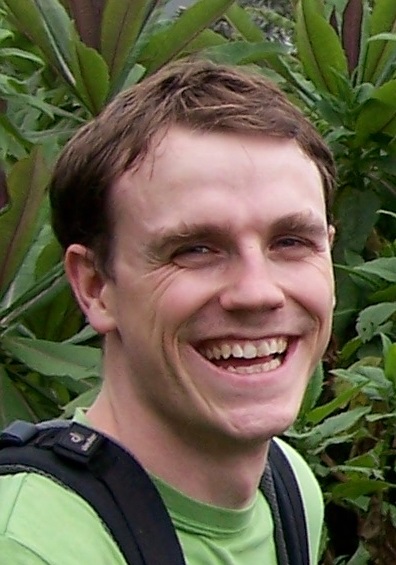}}] {Dirk Witthaut}  
received his Diploma (MSc) and PhD in Physics from the Technical University of Kaiserslautern, Germany, in 2004 and 2007, respectively. He has been a Postdoctoral Researcher at the Niels Bohr Institute in Copenhagen, Denmark and at the Max Planck Institute for Dynamics and Self-Organization in G\"ottingen, Germany and a Guest Lecturer at the Kigali Institute for Science and Technology in Rwanda. Currently, he is leading a Research Group at Forschungszentrum J\"ulich, Germany and he is a junior professor at the University of Cologne.  \end{biography}

\end{document}